\documentclass[english,aps, twocolumn]{revtex4}
\usepackage[T1]{fontenc}
\usepackage[latin9]{inputenc}
\usepackage{float}
\usepackage{units}
\usepackage{amsmath}
\usepackage{amssymb}
\usepackage{graphicx}
\usepackage{esint}

\makeatletter

\providecommand{\tabularnewline}{\\}

\@ifundefined{textcolor}{}
{%
 \definecolor{BLACK}{gray}{0}
 \definecolor{WHITE}{gray}{1}
 \definecolor{RED}{rgb}{1,0,0}
 \definecolor{GREEN}{rgb}{0,1,0}
 \definecolor{BLUE}{rgb}{0,0,1}
 \definecolor{CYAN}{cmyk}{1,0,0,0}
 \definecolor{MAGENTA}{cmyk}{0,1,0,0}
 \definecolor{YELLOW}{cmyk}{0,0,1,0}
}

\newcommand{\tvect}[2]{%
  \ensuremath{\bigl(\negthinspace\begin{smallmatrix}#1\\#2\end{smallmatrix}\bigr)}}

\newcommand{\tmatrix}[4]{%
  \ensuremath{\bigl(\negthinspace\begin{smallmatrix}#1&#2\\#3&#4\end{smallmatrix}\bigr)}}

\makeatother

\usepackage{babel}
\begin{document}

\title{Dynamic Coherence in Excitonic Molecular Complexes \\ under Various
Excitation Conditions}

\author{Aurélia Chenu, Pavel Malý and Tomáš Man\v{c}al}

\address{Faculty of Mathematics and Physics, Charles University in Prague,
Ke Karlovu 5, 121 16 Prague 2, Czech Republic}
\begin{abstract}
We investigate the relevance of dynamic quantum coherence in the energy
transfer efficiency of molecular aggregates. We contrast the dynamics
after excitation of a quantum mechanical system with that of a classical
system. We demonstrate how a classical description of an ensemble
average can be satisfactorily interpreted either as a single system
driven by a continuous force or as an ensemble of systems each driven
by an impulsive force. We derive the time evolution of the density
matrix for an open quantum system excited by light or by a neighboring
antenna. We argue that unlike in the classical case, the quantum description
does not allow for a formal decomposition of the dynamics into sudden
jumps in the quantum mechanical state. Rather, there is a natural
finite time-scale associated with the excitation process. We propose
a simple experiment how to test the influence of this time scale on
the yield of photosynthesis. Because photosynthesis is intrinsically
an average process, the efficiency of photosynthesis can be assessed
from the quantum mechanical expectation value calculated from the
second-order response theory, which has the same validity as the perturbative
description of ultrafast experiments. We demonstrate using typical
parameters of the currently most studied photosynthetic antenna, the
Fenna-Matthews-Olson (FMO) complex, and a typical energy transfer
rate from the chlorosome baseplate, that dynamic coherences are averaged
out in the complex despite excitation proceeding through a coherent
superposition of its eigenstates. The dynamic coherence averages out
even when the FMO model is completely free of all dissipation and
dephasing. We conclude that under natural excitation conditions\emph{,
}\textsl{\emph{coherent dynamics cannot be responsible for the remarkable
efficiency of the photosynthesis}}\emph{,} even when considering the
dynamics at a single molecular level. 
\end{abstract}
\maketitle

\section{Introduction\label{sec:Introduction}}

In the light-harvesting antennae of both plants and bacteria and in
other molecular aggregates and crystals, electronic excited state
delocalization has been since long time ago known to play a significant
role in establishing their energy transport properties (e.g. \cite{Foerster1965a,Knox1975a,Scholes2003a,Jang2008a}).
Delocalization of the electronic eigenstates over more than one pigment
corresponds to a correlation between electrons of different pigments
in the same molecular crystal or aggregate which results from their
direct mutual electrostatic interaction. Correspondingly, this effect
has been often referred to as\emph{ electronic quantum coherence.}
In photosynthetic antennae, the excited eigenstate delocalization
enables fast transfer of excitation in space and, in combination with
the influence of the virtually infinite number of degrees of freedom
(DOF) of the protein environment, directs the energy flow in a funnel
like fashion towards the energetically lower lying states of the photosynthetic
reaction center (RC). Here the chemical part of the photosynthesis
starts. The several order of magnitude difference in the energy transfer
rates and the rates of the excitation recombination in the antenna
(inverse picoseconds vs. inverse nanoseconds, respectively) explains
then the remarkable efficiency of the photosynthetic antennae. Because
light interacts resonantly with the electronic excited states, and
because the wavelength of the light is usually several orders of magnitude
larger than the photosynthetic antennae, the light excites the delocalized
eigenstates rather than the individual molecules themselves. This
invites a kinetic picture in which the antenna is represented by a
set of excited states with energy relaxation between them. The ultra
short laser light pulses used in laboratory experiments inevitably
excite superpositions of these antenna eigenstates, because they spectrally
cover several of them. Nevertheless, the kinetic picture usually assumes
independence of the oscillating coherent components of the corresponding
electronic density matrix from the much slower varying population
components. The presence of any coherence induced by lasers has thus
no bearing on the energy transfer in this picture. This assumption
is usually termed secular approximation.

With the advent of the two-dimensional (2D) coherent Fourier transformed
spectroscopy \cite{Brixner2004b,Cowan2004a,Brixner2004a}, a technique
became available which could detect the time-dependent signatures
of the superpositions of the antenna excited eigenstates. The properties
of the oscillating signatures were predicted for the photosynthetic
Fenna-Matthews-Olson (FMO) pigment-protein complex (PPC) \cite{Pisliakov2006a},
and the oscillations were soon observed experimentally \cite{Engel2007a}.
Similar oscillations were apparently observed even earlier on the
same system in Ref. \cite{Prokhorenko2002a}, but it was only with
the availability of the new spectroscopic technique that the effect
received widespread attention. Similar oscillations have since been
observed in other photosynthetic complexes such as the RC of a purple
bacteria \cite{Lee2007a}, the light-harvesting complex II (LHCII)
from higher plants \cite{Calhoun2009a,Mercer2009a}, from marine algae
\cite{Mercer2009a,Collini2010a} and from purple bacteria \cite{Harel2012a}
at room temperature (RT), as well as in conjugated polymers \cite{Collini2009a}.
Somewhat unfortunately, this transient oscillatory effect was consistently
referred to as \emph{electronic quantum coherence}, i.e. with the
term previously assigned to excited state delocalization. In order
to lift the overloading of the term quantum coherence, we will refer
throughout this paper only to the quantum beats observed by the 2D
spectroscopy as quantum coherence. Excited state delocalization is
a well established, almost self explanatory term, and it thus justifies
dropping the earlier meaning of the term quantum coherence. Dynamic
coherence does not require delocalization, nor vice versa. In general,
these two effects are entirely unrelated. Delocalization has, however,
an effect on the life-time of the dynamic coherence, once it is created
in the system. A recent study which clearly distinguishes the different
phenomena currently termed quantum coherence in the literature and
argues that only exciton delocalization can have an effect on energy
transfer is e.g. Ref. \cite{Kassal2013a}.

One remarkable deviation from the early predictions about the coherent
oscillations in FMO was their unexpectedly long life-time, which reaches
beyond 1 ps. It was immediately speculated that the surprisingly long
life-time of the oscillations reflects some intrinsic dynamic quantum
mechanical behavior of the photosynthetic antennae, which is crucial
in achieving their high efficiency. Table \ref{tab:experiment} presents
the measured life-times of the observed quantum beats in the different
photosynthetic complexes. These oscillations have been mostly interpreted
as a signature of superpositions of electronic eigenstates. While
in some cases, their life-time is only few hundreds of femtoseconds
\cite{Collini2010a,Harel2012a,Turner2012a}, several photosynthetic
complexes exhibit surprisingly long-lived oscillations \cite{Engel2007a,Lee2007a,Calhoun2009a,Mercer2009a,Panitchayangkoon2010a,Panitchayangkoon2011a,Turner2012a}.
A variety of theoretical models assuming an electronic origin have
been considered, and despite the diversity of approaches, they predict
oscillations with similarly short life-time, i.e. some hundreds of
fs (see \cite{Pachon2011a}, references therein, and also \cite{Kreisbeck2012a,Hein2012a,Fujita2013a}).
Although it is sometimes argued that this life-time is `long', it
is still well below the ps time scale observed in a number of experiments
at low temperature (e.g. \cite{Panitchayangkoon2011a}), where oscillations
still persist at the end of the measurement time limit. It is important
to note that the expected life-time of the oscillations in FMO, at
the time of their first measurements, was based on the estimate of
Ref. \cite{Pisliakov2006a}, which used the relaxation rates calculated
in Ref. \cite{Cho2005a}. According to Ref. \cite{Pisliakov2006a},
$100-300$ fs is a characteristic value for the coherence life-time,
and for the coherence between e.g. the two lowest energy excitons,
the estimated life-time is $\tau_{12}\approx540$ fs, based on the
rates calculated in Ref. \cite{Cho2005a}. It is with respect to this
life-time that the oscillations measured in Ref. \cite{Engel2007a}
were considered long. In all these theoretical estimates, the effects
of delocalization are fully taken into account, yet the predicted
life-times of coherence are remarkably shorter than the measured values.
This invalidates the argument that the measured dynamic coherence
is somehow an experimental measure of eigenstate delocalization. 

The main mechanism currently suspected to extend the electronic coherence
life-time over a ps assumes spatial correlation between environmental
degrees of freedom (DOF) of different pigments \cite{Lee2007a,Mohseni2008a,Rebentrost2009a,Hayes2010a,Ishizaki2010a,Wu2010a,Abramavicius2011a,Struempfer2011a,Caycedo-Soler2012a,Caram2012a}.
Such correlation was however not confirmed by molecular dynamics calculations
\cite{Olbrich2010a,Shim2012a}. Recent theories based on resonance
coupling between the electronic and vibrational states \cite{Christensson2012a,Tiwari2013a,Chenu2013a,Chin2013a}
successfully predict picosecond life-time. It should be noticed that
this model, by considering explicitely vibrational DOF of the molecular
system, effectively introduces correlation between the bath of different
states \textit{on the same pigment}. As such, it is in line with the
assumption of correlated environmental DOF without contradiction with
results from molecular dynamics simulations.

The observation of quantum beating in various complexes seems to suggest
that this is a general phenomenon in photosynthetic systems under
the laboratory excitation conditions. But whether these effects are
of importance for the high efficiency of natural photosynthesis or
whether they are just artifacts of the ultra fast excitation used
in laboratory situation remains a disputed question. As inferred by
the authors from the current debate, it is often assumed that the
dynamic coherence plays an essential role in light-harvesting efficiency.
Such an assumption rests on the notion that a time dependent effect,
similar to the one observed in time-resolved experiments with ultra
fast pulses, also accompanies light-harvesting of natural sunlight.
Suggestions have been made that a thermal source consists of a collection
of random femtosecond pulses, and that the experimental observations
would be representative of the \textsl{in vivo} conditions \cite{Cheng2009a}.
However, based on both semi-classical and fully quantum mechanical
representation of the light-matter interaction, it has been demonstrated
that excitation by incoherent light as well as one-photon absorption
lead to a stationary mixed state, and cannot create a superposition
of energy eigenstates (i.e. dynamic coherence), neither on isolated
molecules \cite{Jiang1991a,Brumer2012a} nor on open systems \cite{Mancal2010a,Pachon2012b}.
Thereby, the existence of quantum coherence has been shown to be dependent
on the excitation process \cite{Mancal2010a,Brumer2012a}, which is
in contrast with a qualitative description suggesting that the absorption
of a single photon triggers the same coherent molecular response,
regardless of the character of the light source \cite{Ishizaki2011a}.

We are apparently left with two pictures supposedly explaining efficiency
of the EET in photosynthesis. The earlier kinetic picture in which
coherent oscillations, although present, have no influence on the
energy transfer, and the newer coherent picture, in which coherence
is posited to have a crucial role. It is thus either the interplay
of delocalization with the environment or the interplay of quantum
coherence (in the above defined sense) with the environment which
lead to high efficiency of the photosynthesis. One might suggest a
combined picture, in which the effect of coherence plays a role of
a correction to the kinetic picture. In this case the influence of
coherence would occur through the coupling between the oscillating
terms of the density matrix and the eigenstate population elements,
going thus beyond the secular approximation. Such coupling was indeed
observed in Ref. \cite{Panitchayangkoon2011a}. The relative contribution
of this effect to the EET dynamics remains a question. Predictions
using the secular approximation already lead to a close to unity efficiency
of the energy transfer, and it is therefore questionable that the
improvement on the order of several per cent would play a role for
natural photosynthesis, especially considering that it is often even
required to decrease the efficiency drastically to avoid damage by
over-excitation. There is so-far no evidence that non-secular effects
play any significant role in the EET in photosynthetic antennae.

The aim of this paper is to clarify the relation between the dynamics
induced by an ultrafast excitation and the one corresponding to a
slow driving force. In particular we will attempt to shed light on
the time-scale on which excitation of a molecular system occurs when
it is driven by an external source of excitation, such as light or
a neighboring antenna. In Section \ref{sec:Relation-between-Impulsive}
we will discuss certain equivalence between a slowly driven dynamics
of a single classical system and an average over time-evolution of
an ensemble of classical systems excited impulsively. In Section \ref{sec:Excitation-by-Light}
we formulate the theory of the excitation the molecular system by
light, and in Section \ref{sec:Excitation-by-Neighboring} we use
the same formulation to describe an excitation transfer from another
molecular system. We discuss the time-scale of the excitation events
in Section \ref{sec:Time-Scale-of-Excitation} and we propose a simple
experiment which could test the influence of this time scale on the
yield of photosynthesis. We demonstrate the properties of the excited
state of a model antenna under excitation feeding from a neighboring
system in Section \ref{sec:Demo}. Our conclusions are presented in
Section \ref{sec:Conclusions}.

\begin{widetext}

\begin{table}[h]
\begin{tabular}{|c|c|c|cr|}
\hline 
Measured system  & Exp. tech.  & Temp.  & Life time  & \tabularnewline
\hline 
\hline 
FMO (\textsl{Chlorobium tepidum}) & 2D ES & 77 K  & $>$660 fs  & \cite{Engel2007a}\tabularnewline
\hline 
 &  & 277 K  & $>$300 fs  & \cite{Panitchayangkoon2010a}\tabularnewline
\hline 
 &  & 77 K  & $>$1800 fs  & \cite{Panitchayangkoon2011a}\tabularnewline
\hline 
RC (\emph{Rhodobacter Sphaeroides})  & 2CECPE & 77 K  & 440 fs  & \cite{Lee2007a}\tabularnewline
\hline 
 &  & 180 K  & 310 fs  & \cite{Lee2007a}\tabularnewline
\hline 
LH2 (\emph{Rhodobacter Sphaeroides})  & 2D ES  & RT  & 400 fs  & \cite{Harel2012a}\tabularnewline
\hline 
LH2 (purple bacteria)  & ARC WM  & RT  & > 2800 fs  & \cite{Mercer2009a}\tabularnewline
\hline 
LHCII (\emph{Arabidopsis Thaliana$)$}  & 2D ES  & 77 K  & $>$500 fs  & \cite{Calhoun2009a}\tabularnewline
\hline 
PC645 and PE545 (marine cryptophytes)  & 2D PES  & 294 K  & $>$130 fs  & \cite{Collini2010a,Turner2012a}\tabularnewline
\hline 
\end{tabular}

\caption{Characteristics of the quantum beats observed in various photosynthetic
systems by different experimental techniques: 2D Electronic Spectroscopy
(2DES), 2D Photo Echo Spectroscopy (2DPES), Two-Color Electronic Coherence
Photon Echo (2CECPE) and angle-resolved coherent wave mixing (ARC
WC). The sign ''>'' means that the coherence life-time was estimated
to be larger than the given value.}

\label{tab:experiment} 
\end{table}

\end{widetext}

\section{Relation between Impulsive and Slowly Driven Dynamics in Classical
Systems\label{sec:Relation-between-Impulsive}}

The relation between the dynamics observed in an ultrafast excitation
by a pulsed laser and the one which occurs when the system is driven
by a weak steady illumination can be illustrated on a completely classical
example of driving the dynamics of a classical system. Even in a laboratory
experiment with laser pulses, the excited state dynamics of molecular
systems relevant for photosynthesis can be described by introducing
a source term into the reduced density matrix equations. In the density
matrix formalism, this term is linear in excitation field intensity
(i.e. quadratic in electric field). Validity of the description by
such linear term in intensity is the same as the third order perturbation
theory description of the four-wave mixing (FWM) experiments, such
as the 2D Fourier transformed spectroscopy. After a molecular system
is excited in a FWM experiment by two elementary light-matter interactions
(2$^{nd}$ order), there is only one more excitation field acting
linearly and producing the stimulated signal \cite{MukamelBook}.
This signal can be experimentally verified to be of the third order
in field. \textsl{Second order treatment of the laboratory and Sun
light excitation is therefore completely satisfactory.} When treated
in a wavefunction formalism, the source term can even be linear in
the exciting field, because the calculated observable is quadratic
in the wavefunction.

To illustrate the relation between the dynamics of a system driven
by a short external impulsive force and a system driven by a slowly
varying force, let us consider a classical oscillator which is driven
by an external force $f(t)$. The corresponding equations of motion
read
\begin{equation}
\begin{array}{c}
\dot{p}=-\omega\, q+f(t)\\
\dot{q}=+\omega\, p.
\end{array}
\end{equation}
This problem can be rewritten formally into the form
\begin{equation}
\dot{\phi}(t)-H\,\phi(t)=\varphi_{0}\, f(t),\label{eq:full_eq}
\end{equation}
where $\phi(t)=\tvect{p}{q}$ represents the phase-space vector consisting
of the oscillator coordinate and its conjugated momentum, $\varphi_0=\tvect{0}{1}$
ensures that $f(t)$ drives only the momentum, and H=\tmatrix{0}{+\omega}{-\omega}{0}.
The solution of the full equation, Eq. (\ref{eq:full_eq}), can be
found by first finding the solution of the same equation, but with
an ``ultrashort'' driving term on its right hand side (r.h.s.),
so-called particular solution $\phi_{\delta}(t)$. For this we have
\begin{equation}
\dot{\phi}_{\delta}(t)-H\,\phi_{\delta}(t)=\varphi_{0}\,\delta(t).\label{eq:delta_eq}
\end{equation}
The particular solution of this equation can be written in form of
a Green's function $G(t)$ as
\begin{equation}
\phi_{\delta}(t)=G(t)\varphi_{0}=\Theta(t)\left(\begin{array}{c}
\sin\omega t\\
\cos\omega t
\end{array}\right),\label{eq:delta_sol}
\end{equation}
where
\begin{equation}
G(t)=\Theta(t){\rm exp}\{Ht\}=\Theta(t)\left(\begin{array}{cc}
\cos\omega t & \sin\omega t\\
-\sin\omega t & \cos\omega t
\end{array}\right),
\end{equation}
and $\Theta(t)$ is the Heaviside step function. A reaction of the
oscillator to the ``shock-like'' ultrashort excitation is an oscillatory
motion. In case of a general driving force, the evolution of the phase-space
vector can be shown to be given by 
\begin{equation}
\phi(t)=\int\limits _{0}^{\infty}{\rm d}\tau\, G(t-\tau)\varphi_{0}\, f(\tau),\label{eq:sol_general}
\end{equation}
which indicates that the dynamics of a system under any type of excitation
can be found by averaging in time over the transient impulsively excited
dynamics with the weight given by the driving function $f$. This
includes even the situation when the driving force is constant, $f(t)=f_{0}$.
In this case, the solution can be found intuitively to be a stationary
displaced oscillator 
\begin{equation}
\phi_{f_{0}}(t)=\left(\begin{array}{c}
\frac{f_{0}}{\omega}\\
0
\end{array}\right).\label{eq:sol_stac}
\end{equation}
 The same result can be obtained from Eq. (\ref{eq:sol_general})
by assuming that sometime in the past, the function $f(t)$ was slowly
switched on in such a way that $\frac{d}{dt}f(t)\approx0$ at all
times. The function became constant, $f(t)=f_{0}$, at some later
time. Substituting the definitions of all quantities into Eq. (\ref{eq:sol_general}),
using $\tau^{\prime}=t-\tau$ and integrating by parts, we get
\begin{align} \label{eq:slow_sol}
\phi(t) & =\int\limits _{0}^{t}{\rm d}\tau^{\prime}\left(\begin{array}{c}
\sin\omega\tau^{\prime}\\
\cos\omega\tau^{\prime}
\end{array}\right)f(t-\tau^{\prime}) \\
 & =\frac{f(t)}{\omega}\left(\begin{array}{c}
1\\
0
\end{array}\right)-\frac{1}{\omega}\int\limits _{0}^{t}{\rm d}\tau\left(\begin{array}{c}
-\cos\omega\tau^{\prime} \\
\sin\omega\tau^{\prime}
\end{array}\right)\frac{d}{d\tau}f(t-\tau)\nonumber.
\end{align}
The second term on the last r.h.s. of Eq. (\ref{eq:slow_sol}) can
be made arbitrarily small by making the switching-on process arbitrarily
slow. The first term corresponds to Eq. (\ref{eq:sol_stac}), because
$f(t)=f_{0}$ after the transient switching on regime is over. It
is thus clear that \textsl{the stationary solution}, Eq. (\ref{eq:sol_stac}),
\textsl{can be described as an interference of contributions created
by ultrashort impulses}. 

When an ensemble of $N$ classical systems is treated, we might be
interested in an ensemble value of some quantity, e.g. 
\begin{equation}
\Phi(t)=\frac{1}{N}\sum_{n}\phi_{n}(t)=\langle\phi(t)\rangle.\label{eq:ave_Psi}
\end{equation}
Here $\phi_{n}(t)$ is the phase vector of an individual member of
the ensemble, which might be in principle assigned its own Green's
function $G_{n}(t)$ and initial condition $\varphi_{0}^{(n)}$. If
we assume these to be identical for all members of the ensemble, the
average in Eq. (\ref{eq:ave_Psi}) can be replaced by $\phi(t)$ itself,
i.e. $\Phi(t)=\phi(t)$.

One can reformulate the result, Eq. (\ref{eq:sol_general}), in terms
of a formal virtual ensemble. One can assume that all members of the
ensemble are identical, but the constant driving is in fact not constant
for each individual member of the ensemble, rather, it is constant
only in its effect of the ensemble quantity, so that
\begin{equation}
\Phi(t)=\sum_{n}\int\limits _{0}^{\infty}{\rm d}\tau\ G(t-\tau)\varphi_{0}f_{n}(\tau)\label{eq:Phi_suma}
\end{equation}
with 
\begin{equation}
f(t)=\sum_{n}f_{n}(t),\label{eq:f_in_fn}
\end{equation}
where $f_{n}(t)$ acts only on the $n^{th}$ system of the ensemble.
We can further assume that the force $f_{n}(t)$ acts only during
a very short interval $\Delta t$ after time $t_{n}$, where the times
$t_{n}$ are distributed on the time axis with the step $\Delta t$.
The $f_{n}(t)$ then reads as
\begin{equation}
f_{n}(t)=Nf(t)\Theta(t-t_{n})\Theta(-(t-(t_{n}+\Delta t))),
\end{equation}
where the two Heaviside step functions $\Theta(t)$ ensure that the
function $f_{n}(t)$ is equal to $f(t)$ on the interval $[t_{n};\, t_{n}+\Delta t]$
and zero otherwise. When $\Delta t$ is infinitesimal, Eq. (\ref{eq:f_in_fn})
can be written as 
\begin{equation}
f(t)=\sum_{n}\Delta t\ f(t_{n})\frac{\Theta(t-t_{n})\Theta(-(t-(t_{n}+\Delta t)))}{\Delta t},\label{eq:f_in_ftn}
\end{equation}
where we replaced $f(t)$ by its values $f(t_{n})$ at the beginning
of the interval $[t_{n};\, t_{n}+\Delta t]$ . The sum over $n$ is
meant here as a sum over times at which these molecules are being
driven by the force $f_{n}(t)$. In the limit of $\Delta t\rightarrow0$,
we can write Eq. (\ref{eq:f_in_ftn}) as an integral
\begin{equation}
f(t)=\int\limits _{0}^{\infty}{\rm d}t^{\prime}f(t^{\prime})\delta(t-t^{\prime}).
\end{equation}
 The integral over $t^{\prime}$ could still be understood as an integral
over an ensemble of molecules. The overall vector for the ensemble
is therefore:
\begin{align}
\Phi(t) & =\int\limits _{0}^{\infty}{\rm d}t^{\prime}\int\limits _{0}^{\infty}{\rm d}\tau\ G(t-\tau)\,\varphi_{0}\,\delta(\tau-t^{\prime})f(t^{\prime})\label{eq:Phi_final}\\
 & =\int\limits _{0}^{\infty}{\rm d}t^{\prime}\ G(t-t^{\prime})\varphi_{0}f(t^{\prime})=\phi(t).\nonumber 
\end{align}
The first line of the last equation should be interpreted as a sum
of dynamics of individual systems which have been excited at respective
times $t-t^{\prime}$ in the past and oscillate without relaxation
ever since. The second line highlights the fact that this is equivalent
to the general solution for a single system $\phi(t)$ (\textsl{cf.}
Eq. \ref{eq:sol_general}). 

The equations of motion of a classical driven oscillator represent
therefore at least two very different physical problems: either (i)
a single oscillator driven by a continuous force or (ii) an ensemble
of oscillators driven each by an impulsive force. Because the resulting
dynamics is the same, we can always represent single oscillator by
a virtual ensemble of oscillators, whenever it is advantageous in
the mathematical treatment of the problem. 

Classical dynamics with an instantaneous excitation is only a mathematical
prerequisite for the calculation of an actual dynamics of the system,
which is determined by the driving force. On the other hand, one can
imagine a physical representation of the action of a steady force
$f(t)$ and its decomposition Eq. (\ref{eq:f_in_fn}), say, by a stream
of particles with small cross-section acting on an ensemble of oscillators.
Each collision leads to a transition of momentum from the particle
to the oscillator, and the oscillator follows Eq. (\ref{eq:delta_eq})
with an instantaneous driving force. The average quantity $\Phi(t)$
follows Eq. (\ref{eq:Phi_suma}) or equivalently Eq. (\ref{eq:Phi_final}).
This physical representation of Eq. (\ref{eq:Phi_final}) seems to
form a basis of an intuitive picture of quantized light acting on
an ensemble of molecules. While in the above classical picture it
is quite natural to deal with instantaneous excitation events, we
will see in the following sections that quantum mechanics prevents
us from introducing such ultrafast excitations. The instantaneous
quantum jumps seem to be only a simplified interpretation imported
to quantum mechanics from classical mechanics.

\section{Excitation of an Open Quantum System by Light\label{sec:Excitation-by-Light}}

In this section we will deal with the problem of external driving
from a quantum mechanical perspective. Let us introduce the Hamiltonian
$H$ of a system consisting in a molecular aggregate $M$ interacting
with a radiation $R$, its immediate environment (or bath) $B$ and
their mutual interactions
\begin{equation}
H=H_{R}+H_{B}+H_{M}+H_{M-B}+H_{M-R}.\label{eq:hamiltonian_radiation}
\end{equation}
Here, $H_{M-B}$ is the interaction between the molecular system (or
more precisely its selected DOF, such as electronic states) with the
environment, and $H_{M-R}$ is the interaction of the molecular system
with light. We neglect the interaction of the environment with the
light and assume that the interaction of the molecular subsystem with
the light is weak. 

The time evolution of the system is given by the Schr\"{o}dinger
equation $i\hbar\frac{\partial}{\partial t}|\psi(t)\rangle=H|\psi(t)\rangle$.
Typically, the molecular system is in its electronic ground state
$|g\rangle$ before the light is applied. It is excited by light into
a band of states $|n\rangle$. We define complementary projectors
$P_{g}=|g\rangle\langle g|$ and $P_{e}=\sum_{n}|n\rangle\langle n|$
which define the unity operator $1\equiv P_{g}+P_{e}$ on the relevant
electronic Hilbert space. We ignore all dark states or states that
could be reached by multiple excitation by light (such as two-exciton
states etc.) because we assume weak light-matter interaction which
allows for a linearization in the interaction Hamiltonian $H_{M-R}$.

Motivated by the usual situation in photosynthetic antennae, we assume
that the molecular Hamiltonian $H_{M}$ does not include terms directly
connecting the ground state with the excited states (the transitions
between electronic states separated by optical gap are only mediated
by light). Similarly, the system--bath interaction $H_{M-B}$ does
not cause such transitions on the time-scales similar or shorter than
the excitation energy transfer. These properties are expressed in
the relations
\begin{equation}
P_{g}\, H_{M}\, P_{e}=P_{g}\, H_{M-B}\, P_{e}=0.
\end{equation}
The system--light interaction term, on the other hand, contains just
terms causing transitions between the ground- and excited states and
therefore
\begin{equation}
P_{e}\, H_{M-R}\, P_{e}=0.
\end{equation}
The operators $H_{R}$ and $H_{B}$ do not depend on the electronic
DOF of the molecular subsystem and thus $P_{e}\, H_{R}\, P_{e}=P_{g}\, H_{R}\, P_{g}=H_{R}$,
and similarly for $H_{B}.$ We will define the system--bath interaction
Hamiltonian so that $P_{g}\, H_{M-B}\, P_{g}=0$ as any difference
from zero in $H_{M-B}$ of the ground state can be added to the Hamiltonian
$H_{B}$. With the following definitions 
\begin{equation}
H_{e}\equiv P_{e}H_{M}P_{e},\ H_{g}\equiv P_{g}H_{M}P_{g},\ 
\end{equation}
\begin{equation}
H_{eg}\equiv P_{e}H_{M-R}H_{g},\ \Delta V\equiv P_{e}H_{M-B}P_{e},
\end{equation}
\begin{equation}
|\psi_{g}(t)\rangle\equiv P_{g}|\psi(t)\rangle,\ |\psi_{e}(t)\rangle\equiv P_{e}|\psi(t)\rangle,
\end{equation}
we can write two coupled Schr\"{o}dinger equations for the ground
state and the excited state bands
\begin{equation}
\begin{split}
i\hbar\frac{\partial}{\partial t}|\psi_{e}(t)\rangle=&\left(H_{R}+H_{B}+H_{e}+\Delta V\right)|\psi_{e}(t)\rangle\\
&+H_{eg}|\psi_{g}(t)\rangle,\label{eq:schr_p1}
\end{split}
\end{equation}

\begin{equation}
\begin{split}
i\hbar\frac{\partial}{\partial t}|\psi_{g}(t)\rangle=&\left(H_{R}+H_{B}+H_{g}\right)|\psi_{g}(t)\rangle\\
+&H_{ge}|\psi_{e}(t)\rangle.\label{eq:schr_p2}
\end{split}
\end{equation}
The last term in Eq. (\ref{eq:schr_p2}) will be neglected as we aim
for an equation for excited state evolution which is linear in $H_{M-R}$.
We also set the electronic energy of the ground state to zero, i.e.
$H_{g}=0$. Given an initial condition $|g\rangle$ at time $t=0$,
we can solve Eq. (\ref{eq:schr_p2}), obtaining 
\begin{equation}
|\psi_{g}(t)\rangle=U_{R}(t)U_{B}(t)|g\rangle,\label{eq:solve_psi_g}
\end{equation}
where we have defined the evolution operators $U_{R}(t)=\exp\{-iH_{R}t/\hbar\}$
and $U_{B}(t)=\exp\{-iH_{B}t/\hbar\}$. 

The solution of Eq. (\ref{eq:schr_p1}) can now be found by using
the solution for the ground state, Eq. (\ref{eq:solve_psi_g}), and
finding the Green's function of its homogenous part\textbf{
\begin{equation}
G(t)=\Theta(t)U_{R}(t)\tilde{G}(t)U_{B}(t),
\end{equation}
}where we defined 
\begin{equation}
\tilde{G}(t)=\exp_{\leftarrow}\left\{ -\frac{i}{\hbar}\left(H_{e}t+\int_{0}^{t}{\rm d}\tau\:\Delta V(-\tau)\right)\right\} ,
\end{equation}
 $\Delta V(t)=U_{B}^{\dagger}(t)\,\Delta V\, U_{B}(t)$ and $H_{eg}(t)=U_{R}^{\dagger}(t)\, H_{eg}\, U_{R}(t)$.\textbf{
}We obtain
\begin{equation}
|\psi_{e}(t)\rangle=U_{R}(t)\int\limits _{0}^{t}{\rm d}\tau\ \tilde{G}(t-\tau)\, H_{eg}(t)\, U_{B}(t)\,|g\rangle.\label{eq:Psi_e_final}
\end{equation}

From Eq. (\ref{eq:Psi_e_final}), we can easily construct the density
operator describing the excited state part of the total system
\begin{equation}
\hat{W}(t)=|\psi_{e}(t)\rangle\langle\psi_{e}(t)|.
\end{equation}
The corresponding reduced density matrix (RDM) of the molecular system
is obtained by averaging quantum mechanically over the radiational
and the environmental DOF: $\hat{\rho}(t)={\rm Tr}_{R}\left[{\rm Tr}_{B}\{\hat{W}(t)\}\right]$.
The system-light interaction Hamiltonian will be assumed in dipole
approximation
\begin{equation}
H_{eg}(t)=-\hat{\mu}_{eg}\,{\cal \hat{{\cal E}}}^{(+)}(t),
\end{equation}
where $\hat{\mu}_{eg}=P_{e}\hat{\,\mu}\, P_{g}$ is a projection of
the transition dipole moment operator, and ${\cal \hat{{\cal E}}}^{(+)}$
is the component of the electric field operator containing the field
annihilation operator. The RDM can now be written as 
\begin{equation}
\hat{\rho}(t)=-\int\limits _{0}^{t}{\rm d}\tau\int\limits _{0}^{t}{\rm d}\tau^{\prime}R_{{\rm light}}(t,\tau,\tau^{\prime})I_{{\rm light}}(\tau,\tau^{\prime}),\label{eq:state_final}
\end{equation}
with 
\begin{equation}
\begin{split}
R_{{\rm light}}(t,\tau,\tau^{\prime})=\left(\frac{i}{\hbar}\right)^{2}{\rm Tr}_{B}\Big\{\tilde{G}(t-\tau)\hat{\mu}_{eg}\\
\times U_{B}(t)|\phi_{g}\rangle\langle\phi_{g}|U_{B}^{\dagger}(t)\hat{\mu}_{ge}\tilde{G}^{\dagger}(t-\tau^{\prime})\Big\},
\end{split}
\end{equation}
\begin{equation}
I_{{\rm light}}(\tau^{\prime},\tau)={\rm tr}_{R}\left\{ {\cal \hat{{\cal E}}}^{(-)}(\tau^{\prime}){\cal \hat{{\cal E}}}^{(+)}(\tau)|\Xi\rangle\langle\Xi|\right\} .\label{eq:EE_corr}
\end{equation}
This is a convenient version of the result from Ref. \cite{Mancal2010a}
where we assumed $|\psi_{0}\rangle=|g\rangle|\varphi_{g}\rangle|\Xi\rangle$,
with $|\Xi\rangle$ representing the state of the light at $t=0$
and $|\varphi_{g}\rangle$ representing the state of the environment.
The operator $R_{{\rm light}}(t,\tau,\tau^{\prime})$ describes the
molecular system response to the action of weak light, which is described
by the field correlation function $I_{{\rm light}}(\tau,\tau^{\prime})$. 

Considering quantized light and the Bose-Einstein distribution of
the modes populations $\bar{n}(\omega)=\nicefrac{1}{(e^{\beta\hbar\omega}-1)}$,
we can arrive at a convenient expression for the correlation function
of light 

\begin{equation}
I(\tau^{\prime},\tau)=\int_{0}^{\infty}d\omega\frac{\hbar}{\varepsilon_{0}2\pi^{2}c^{3}}\frac{\omega^{3}}{e^{\beta\hbar\omega}-1}e^{i\omega(\tau^{\prime}-\tau)},
\end{equation}
where the integrated expression without the exponential is the Planck's
law, denoted $W(\omega)$ in the following. We thus obtain that the
correlation function is stationary, i.e. only function of the difference
$(\tau-\tau')$ and it can be expressed as a Fourier transform of
the spectrum of the light,

\begin{equation}
{\cal G}(\tau)\equiv I(\tau^{\prime},\tau)=\int_{0}^{\infty}d\omega\, W(\omega)\, e^{i\omega\tau}.\label{eq:W-K}
\end{equation}
This formula holds generally for stationary light and is called Wiener-Khintchine
theorem \cite{Loudon2000}. Note the importance of Eq. (\ref{eq:W-K})
which gives the possibility to estimate the light coherence time directly
from its spectrum $W(\omega)$. Also, this allows us the write Eq.
(\ref{eq:state_final}) in a form
\begin{align}
\hat{\rho}(t)=&-\int\limits _{0}^{\infty}{\rm d}\omega\, W(\omega)\\
&\times\left(\int\limits _{0}^{t}{\rm d}\tau\int\limits _{0}^{\tau}{\rm d}\tau^{\prime}R_{{\rm light}}(t,\tau,\tau^{\prime})e^{i\omega(\tau^{\prime}-\tau)}+\textrm{\textrm{h.c.}}\right),\nonumber
\end{align}
using the fact that $R_{{\rm light}}(t,\tau,\tau^{\prime})$ is a
Hermitian operator and integrating separately the parts where $\tau>\tau^{\prime}$
and $\tau<\tau^{\prime}$, respectively. The abbreviation $\textrm{h.c.}$
represents the Hermitian conjugated term. In the case that the driven
system consists of a single excited state $|a\rangle$, the response
takes a form
\begin{align}
\rho_{aa}(t)=&\frac{|\langle a|\hat{\mu}_{eg}|g\rangle|^{2}}{\hbar^{2}}\\
\times&2\,{\rm Re}\int\limits _{0}^{\infty}{\rm d}\omega\, W(\omega) \int\limits _{0}^{t}{\rm d}\tau\int\limits _{0}^{\tau}{\rm d\tau^{\prime}e^{i\omega\tau^{\prime}}e^{-g(\tau^{\prime})-i\omega_{ag}\tau^{\prime}}}, \nonumber
\end{align}
 where we can identify the absorption spectrum
\begin{equation}
\alpha(\omega)\approx|\langle a|\hat{\mu}_{eg}|g\rangle|^{2}2\,{\rm Re}\int\limits _{0}^{\infty}{\rm d}\tau\ e^{i\omega\tau}e^{-g(\tau)-i\omega_{ag}\tau},
\end{equation}
and $g(t)$ the so-called line shape function (Eq. (8.13) in \cite{MukamelBook}).
For times $t$ long relative to the decay time of the linear response,
i.e. after $e^{-g(t)}\approx0$, we get
\begin{equation}
\rho_{aa}(t)\approx\int{\rm d}\omega\, W(\omega)\,\alpha(\omega).\label{eq:tWalpha}
\end{equation}
The population is thus proportional to the overlap of the light-spectrum
and the absorption spectrum of the molecular system, which is an expected
result.

\section{Excitation by Neighboring Antenna\label{sec:Excitation-by-Neighboring} }

Let us now briefly show that Eq. (\ref{eq:state_final}) describes
also another important situation in photosynthesis, namely, the situation
when the relevant molecular system is excited by a neighboring antenna.
For this, we only have to replace the light radiation by an antenna
-- ($R\rightarrow A$) in Eq. (\ref{eq:hamiltonian_radiation}). More
specifically, we replace the Hamiltonian $H_{R}$ of the radiation
by the Hamiltonian of the neighboring antenna
\begin{equation}
H_{A}=H_{B_{A}}+\sum_{k_{A,}l_{A}}H_{k_{A}l_{A}}|k_{A}\rangle\langle l_{A}|,
\end{equation}
where $|k_{A}\rangle$ and $|l_{A}\rangle$ are the electronic excited
states of the neighboring antenna, $B_{A}$ denotes the environment
of the antenna, $H_{B_{A}}$ is its associated Hamiltonian and $H_{k_{A}l_{A}}$
describes the interaction of the antenna electronic states with its
environment. These can also cause transitions between excited states.
The details of the Hamiltonian $H_{A}$ are not of crucial importance
here. The role of the system--light interaction operator is now taken
by the resonance interaction Hamiltonian
\begin{equation}
H_{M-A}=\sum_{k_{A}}\left(\sum_{n}J_{nk_{A}}|n\rangle\langle g|\right)\left(|g_{A}\rangle\langle k_{A}|\right)+\textrm{h.c.}.
\end{equation}
Here, $J_{nk_{A}}$ is the resonant interaction energy between the
state $|k_{A}\rangle$ of the neighboring antenna and the state $|n\rangle$
of the molecular system of interest. The Hamiltonian $H_{M-A}$ has
a somewhat more complicated structure than the light-matter interaction
Hamiltonian and it cannot be in general written as a product of two
operators, each belonging to the Hilbert space of one of the interacting
entities. This complicates the matter only slightly. For simplicity,
let us assume that the excitation is transferred only to one state
$|m\rangle$ of the molecular system. Then the field operator ${\cal \hat{{\cal E}}}{}^{(+)}$
in Eq. (\ref{eq:EE_corr}) will be replaced with the operator ${\cal \hat{A}}^{(+)}=\sum_{k_{A}}J_{mk_{A}}|g_{A}\rangle\langle k_{A}|$
and the operator $\hat{\mu}_{eg}$ simply by $\hat{p}_{eg}=|m\rangle\langle g|$.
We obtain an operator for the molecular response $R_{{\rm antenna}}(t,\tau,\tau^{\prime})$
in which $\hat{\mu}_{eg}$ is replaced by $\hat{p}_{eg}$ and instead
of the light correlation function we have an antenna energy gap correlation
function
\begin{equation}
I_{{\rm antenna}}(\tau,\tau^{\prime})={\rm Tr}_{B_{A}}\left\{ \hat{{\cal A}}^{(-)}(\tau^{\prime})\hat{{\cal A}}^{(+)}(\tau)|\phi_{A}(0)\rangle\langle\phi_{A}(0)|\right\} .
\end{equation}
The state $|\phi_{A}(0)\rangle=|\phi_{M_{A}}(0)\rangle|\phi_{B_{A}}(0)\rangle$
represents the initial state of the antenna (composed of a molecular
aggregate and a bath) at $t=0$, and the trace is taken over the antenna
environment.

If in addition, we assume the neighboring antenna to be represented
by a single level $|l\rangle\equiv|l_{A}\rangle$, it is easy to show
that the expression Eq. (\ref{eq:state_final}) with $I_{{\rm antenna}}$
and $R_{{\rm antenna}}$ represents a F\"{o}rster type energy transfer.
Under these approximations, the response reads
\begin{equation}
R_{{\rm antenna}}(t,\tau,\tau^{\prime})=\frac{1}{\hbar^{2}}{\rm Tr}_{B}\Big\{\tilde{G}_{m}(\tau^{\prime}-\tau)\hat{W}_{B}\Big\}=\frac{1}{\hbar^{2}}e^{-g(\tau^{\prime}-\tau)},
\end{equation}
where we evaluated the trace over the bath by second cumulant expansion
\cite{ValkunasAbramaviciusMancalBook}. The antenna correlation function
reads
\begin{equation}
\begin{split}
I_{{\rm antenna}}(\tau,\tau^{\prime})=|J_{lm}|^{2}{\rm \, Tr}_{B_{A}}\Big\{ U_{A}^{\dagger}(\tau^{\prime}-\tau)|l\rangle\langle g_{A}| \\
\times U_{A}(\tau^{\prime}-\tau)|g_{A}\rangle\langle l|\hat{W}_{B_{A}}\Big\}\rho_{ll}^{(A)}(0),\label{eq:I_antenna}
\end{split}
\end{equation}
where $\rho_{ll}^{(A)}(0)=\textrm{Tr}{}_{B_{A}}\{|\phi_{M_{A}}(0)\rangle\langle\phi_{M_{A}}(0)|\}$
is the population of the level $|l\rangle$, $\hat{W}_{B_{A}}=|\phi_{B_{A}}(0)\rangle\langle\phi_{B_{A}}(0)|$
is the density operator of the antenna environment at initial time,
and $U_{A}(t)=\exp\{-\frac{i}{\hbar}H_{A}t\}$ is the evolution operator
with the Hamiltonian of the antenna. Because we limited the relevant
states of the antenna on a single state $|l\rangle$ and assumed that
it is initially excited, i.e. $|\phi_{M_{A}}(0)\rangle=|l\rangle\Leftrightarrow\rho_{ll}^{(A)}(0)=1$$ $,
Eq. (\ref{eq:I_antenna}) can also be evaluated by cumulant expansion
-- note that the evaluation needs a special care because the antenna
is initially in the excited state. Similar problem is solved when
describing spectrum of spontaneous emission (see e.g. Chap. 8 in \cite{MukamelBook})
and the solution reads
\begin{equation}
I_{{\rm antenna}}(\tau,\tau^{\prime})=|J_{lm}|^{2}e^{-g_{A}^{*}(\tau^{\prime}-\tau)-2i\lambda_{A}(\tau^{\prime}-\tau)+i\omega_{lm}(\tau^{\prime}-\tau)}.
\end{equation}
In case of a single acceptor and a single donor states, Eq. (\ref{eq:state_final})
thus yields
\begin{equation}
\begin{split}
&\rho_{mm}(t)=\frac{2|J_{lm}|^{2}}{\hbar^{2}}\\
&\text{\ensuremath{\times}}{\rm Re}\int\limits _{0}^{t}{\rm d}\tau\int\limits _{0}^{\tau}{\rm d}\tau^{\prime}e^{-g(\tau)-g_{A}^{*}(\tau)-2i\lambda_{A}\tau-i\omega_{ml}\tau}\rho_{ll}^{(A)}(0),\label{eq:rho_nn}
\end{split}
\end{equation}
where we split the double integral from $0$ to $t$ into parts with
$\tau>\tau^{\prime}$ and $\tau<\tau^{\prime}$ and used the property
of the line shape function $g(-t)=g^{*}(t)$. Taking a time derivative
of Eq. (\ref{eq:rho_nn}) we obtain a rate equation with a time dependent
rate 
\begin{equation}
K_{ml}(t)=\frac{2|J_{lm}|^{2}}{\hbar^{2}}{\rm Re}\int\limits _{0}^{t}{\rm d}\tau\ e^{-g(\tau)-g_{A}^{*}(\tau)-2i\lambda_{A}\tau-i\omega_{ml}\tau}.\label{eq:K_tdp}
\end{equation}
For times $t$ when the integrand of Eq. (\ref{eq:K_tdp}) decays
to zero, we can replace the upper limit of the integral by $\infty$.
We have discussed the relation between the correlation function of
the light and its spectrum in the previous section. In the same general
way we can show that the rate given by Eq. (\ref{eq:K_tdp}), for
$t\rightarrow\infty$, has a form of the overlap of the absorption
spectrum of the excitation acceptor and the emission spectrum of the
donor. We have thus obtained an expression which reduces to the standard
F\"{o}rster resonant energy rate. For a general case of energy transfer
between two multilevel systems, Eq. (\ref{eq:state_final}) is equivalent
to the corresponding generalized F\"{o}rster rate description. This
demonstrates the validity and generality of Eq. (\ref{eq:state_final})
for weakly interacting systems. \textsl{\emph{Eq. (\ref{eq:state_final})
therefore provides us with a general platform for discussing the possibility
of excitation of dynamic coherence under different conditions and
studying the relevance of dynamic coherence for the dynamics in excited
state.}}

One might raise a question about the validity of Eq. (\ref{eq:state_final})
for the driving by light or neighboring antenna lasting for times
during which the probability of capturing more than one photon per
molecule becomes significant. Obviously, if the driven system were
limited to some small antenna and the excitation remained on this
antenna for long times, the usual single exciton manifold limitation
of the theoretical description would represent a serious drawback.
However, when the driven system is sufficiently large (e.g. containing
other antennae as well as the reaction center and the different chemical
states of the photosynthetic apparatus into which the excitation is
transferred during the photosynthetic process) the probability of
finding two excitations on a particular antenna remains very low during
the whole duration of illumination. The proper description of an antenna
in operation is a steady state description, where all populations
vary only slowly (see e.g. Ref. \cite{Manzano2013a}). We should emphasize
that expressions such as Eqs. (\ref{eq:tWalpha}) and (\ref{eq:rho_nn})
only represent an illustrative example but that the expression given
in Eq. (\ref{eq:state_final}) can be applied to large system and
is not even limited to just the excitonic part of the photosynthetic
apparatus.

\section{Time-Scale of Excitation Events\label{sec:Time-Scale-of-Excitation}}

It was argued in Refs. \cite{Jiang1991a,Hoki2011a,Mancal2010a,Brumer2012a}
that the state $\hat{\rho}(t)$ produced by Sun light (or in fact
by any stationary source) does not contain any dynamic coherences,
and that the dynamic coherence is therefore irrelevant for the natural
photosynthetic processes. Dynamic coherence can only be created by
excitation of sufficiently short duration. The counter argument against
this conclusion is often based on the short correlation time of the
Sun light, and on a picture where Eq. (\ref{eq:state_final}) represents
an ensemble of molecules in which each molecule has a specific time
of photon capture. This is the bullet-like picture of the photon.
The state, Eq. (\ref{eq:state_final}), is then viewed as a statistical
average over molecules individually excited at specific times. In
the following we will show that the bullet-like picture of the photon
is not consistent with the finite correlation time of Sun light. It
is not possible to write the quantum mechanical result, Eq. (\ref{eq:state_final}),
in terms of sudden excitations in the case that both the correlation
function of light or the material response has a finite correlation
time. While such a formal decomposition into sudden excitations was
possible for classical systems, in quantum mechanics it is not possible,
not even formally. Identifying the dependence of the relevant time-scale
of the excitation process on the light spectrum, we suggest a simple
experiment to test the relevance of this time scale for the yield
of photosythesis. 

The dynamics of the state vector in Eq. (\ref{eq:Psi_e_final}) can
certainly be understood as an integral over instantaneous jumps in
the state vector of the molecular system. However, the result of an
experiment is described by an expectation value

\begin{equation}
\langle\hat{A}\rangle\equiv\langle\psi(t)|\hat{A}|\psi(t)\rangle={\rm tr}\{\hat{A}|\psi(t)\rangle\langle\psi(t)|\}\label{eq:expval}
\end{equation}
which is quadratic in the state vector. Two interactions with the
external fields (one from the bra and one from the ket side of the
expression) are needed to produce the excited state population. In
general, these excitations can occur at different times. Let us select
two infinitesimal contributions to the excited state part $\langle n|\psi_{e}(t)\rangle$
of the wavefunction from Eq. (\ref{eq:Psi_e_final}), where $|n\rangle$
is a selected state from the excited state band. For simplicity we
assume that during the time interval of interest, the state $|n\rangle$
does not relax to other states. Let us assume that the excitations
occured at times $t_{1}$ and $t_{2}$, with $t_{2}>t_{1}$. At time
$t_{1}$ the state of the system corresponds to $|\psi_{g}(t_{1})\rangle=U_{B}(t_{1})U_{R}(t_{1})|g\rangle$
(see Eq. (\ref{eq:solve_psi_g})) and an infinitesimal contribution
to the excitated state in Eq. (\ref{eq:Psi_e_final}) reads
\begin{equation}
|\psi_{e}^{(t_{1})}(t_{1})\rangle=-\frac{i}{\hbar}\hat{\mu}_{eg}\,\hat{\mathcal{E}}^{(+)}|\psi_{g}(t_{1})\rangle{\rm d}t.
\end{equation}
An infinitesimal contribution excited at time $t_{2}$ reads 
\begin{equation}
|\psi_{e}^{(t_{2})}(t_{2})\rangle=-\frac{i}{\hbar}\hat{\mu}_{eg}\,\hat{\mathcal{E}}^{(+)}|\psi_{g}(t_{2})\rangle{\rm d}t.
\end{equation}
At time $t_{2}$, however, the contribution excited at $t_{1}$ has
already evolved for time $(t_{2}-t_{1})$ according to the excited
state Hamiltonian $H_{e}$, that is by evolution opertor $U_{e}(t_{2}-t_{1})=\exp\{-iH_{e}(t_{2}-t_{1})/\hbar\}$.
It reads $|\psi^{(t_{1})}(t_{2})\rangle=U_{e}(t_{2}-t_{1})|\psi^{(t_{1})}(t_{1})\rangle$.
The contribution to the population of some excited state $|n\rangle$
at time $t_{2}$ (and times $t>t_{2}$) is obtained by combining these
two infinitesimal contributions 
\begin{equation}
\delta p_{n}(t_{2},t_{1})=2{\rm \, Re}\langle\psi^{(t_{2})}(t_{2})|n\rangle\langle n|\psi^{(t_{1})}(t_{2})\rangle.\label{eq:dp_def}
\end{equation}
The real part is a result of the fact that in Eq. (\ref{eq:Psi_e_final})
we integrate symmetrically along the times $\tau>\tau^{\prime}$ and
times $\tau<\tau^{\prime}$. Eq. (\ref{eq:dp_def}) leads to
\[
\delta p_{n}(t_{2},t_{1})=\frac{2|\mu_{eg}|^{2}}{\hbar^{2}}({\rm d}t)^{2}
\]
\begin{equation}
\times{\rm Re}\left\{ e^{-g_{nn}(t_{2}-t_{1})-i\omega_{ng}(t_{2}-t_{1})}I(t_{2},t_{1})\right\} ,
\end{equation}
where we used cumulant expansion to evaluate the molecular response,
and we used the definition of the light correlation function, Eq.
(\ref{eq:EE_corr}). For stationary light, the correlation function
depends only on the difference $t_{2}-t_{1}$. Two excitations by
the same field occuring in the wavefunction at times separated by
a delay longer than the life-time of the optical coherence created
in the material or the coherence time of light do not contribute together
to the population. For light with long coherence time, detectors with
sufficiently fast response, i.e. high time resolution, can detect
individual excitation events, i.e. photon arrivals, because different
outcomes of the detection are separated from the particular measured
one by decoherence. This explains the presence of individual ``clicks''
of a detector for the light with coherence time longer than the time
resolution of the detector. The posibility of a faster than coherence
time resolution of the detector, however, does not mean that prior
to the measurement, the light is already in a state where arrival
times are decided. On the other hand, it is important to stress that
Sun light has a spectrum broader than the absorption spectra of photosynthetic
systems, and therefore the correlation time of Sun light is shorter
than the optical dephasing time of the antennae. The characteristic
length of the excitation event in photosynthesis is therefore indeed
decided by the light. 

The correlation function of the Sun light has nevertheless a finite
characteristic coherence time which is given, according to Eq. (\ref{eq:W-K}),
by the Fourier transform of its spectrum. The statistical description
of a quantum mechanical system based on instantaneous quantum jumps
in the wavefunction is not appropriate for systems with finite response
time or light with finite coherence time. It is interesting that such
a statistical description is possible for linear classical systems. 

While a decomposition of the light action on a molecular system in
terms of instantaneous jumps is not possible, nothing can prevent
us from following the suggestion of Ref. \cite{Cheng2009a} were it
was suggested to represent the Sun-light fluctuations by a series
of coherent spikes of short duration occurring at some times $t_{k}$,
so that 
\begin{equation}
I(\tau,\tau^{\prime})=\sum_{k}I_{k}(\tau,\tau^{\prime};t_{k})=\sum_{k}{\cal G}_{k}(\tau-\tau^{\prime};t_{k}).\label{eq:I_expansion}
\end{equation}
Obviously, if the correlation function is only rewritten in the new
way and each molecule feels the same set of spikes, then the predictions
of Refs. \cite{Mancal2010a,Brumer2012a} are valid for individual
molecules and no coherence can be generated even on a single molecular
level. This argument is similar to the one put forward in Ref. \cite{Kassal2013a}
because here it also does not alter what is the order of averaging
in Eq. (\ref{eq:expval}). In order for the proposal of Ref. \cite{Cheng2009a}
to be non-trivial we would have to accept an interpretation of Eq.
(\ref{eq:I_expansion}), in which each molecule of the ensemble feels
e.g. just one component of the decomposition. Correspondingly, each
molecule would be subject to some microscopic field state $|\Xi_{k}\rangle$
for which $I_{k}(\tau,\tau^{\prime};t_{k})={\rm Tr}_{R}\{\hat{\mathcal{E}}(\tau)\hat{\mathcal{E}}(\tau^{\prime})|\Xi_{k}\rangle\langle\Xi_{k}|\}$.
The fact that we detect a macroscopic correlation function $I(\tau,\tau^{\prime})$
would then be a result of the fact that our detectors are macroscopic.
The expansion Eq. (\ref{eq:I_expansion}) could in principle enable
us to simulate the finite time scale of the excitation events. Here
we would no more be exciting all molecules by the same averaged external
field, and correspondingly, if the short excitation time aspect of
the excitation process has an influence on the yield of the photosynthetic
process, the sum of contributions of the individual molecules would
not be guaranteed to yield the same result as in the case where all
molecules are excited by one averaged field. This case is however
beyond the validity of Eq. (\ref{eq:state_final}) in which the decomposion
of Eq. (\ref{eq:I_expansion}) does not affect the final macroscopic
state of the system in any way and correspondingly it has no influence
on the yield. 

Interestingly, one can relatively easily test the proposition of Ref.
\cite{Cheng2009a} experimentally, without any use of a complicated
time-resolved spectroscopy apparatus. For this experimental test one
has to be able to measure the macroscopic yield of the light capture
and the transfer throught a given antenna, e.g. in terms of the number
of excitations transferred to the RC. The measurement can be performed
with the Sun light (carefully attenuated in order to prevent the self
regulating mechanisms of photosynthesis to decrease the optimal yield).
This is a situation where Eq. (\ref{eq:W-K}) suggests that individual
excitations occur on a short time scale. Then one can repeat the same
measurement with a monochromator, selecting spectrally narrow parts
of the Sun-light spectrum. Eq. (\ref{eq:W-K}) predicts that in this
case, the time-scale of the interaction with light is much longer.
When the spectrum of the light is narrower than the spectral distance
between electronic levels of the antenna, the time-scale of the excitation
allowed by the light spectrum is already too long to create any coherences
between the two electronic levels. Finally, one can integrate the
yield across the whole Sun spectrum and compare with the total yield
achieved by the total Sun spectrum. 

Eq. (\ref{eq:state_final}) predicts that the yields are the same.
Notice that it is easy to construct a case where the yields were not
the same. If, for instance, the photosynthetic process to be completed
required first an excitation at certain wavelength followed by an
excitation at another wavelegth, then a single spectral component
would reasult in no yield. This has however nothing to do with the
coherent excitation, and this case is in fact beyond the validity
of Eq. (\ref{eq:state_final}) which assumes that we always start
with exciting the ground state of the antenna. Extending the present
formalism to more complicated initial state of excitation is however
possible.

\section{Dynamic Coherence and Delocalization under Various Excitation Conditions\label{sec:Demo}}

The experiment suggested in the previous section could in principle
decide the question raised by the detection of long lived coherence
in photosynthetic antenna. On the other hand, anticipating its result,
we have every reason to believe that the quantum mechanical description
represented by Eq. (\ref{eq:state_final}) is valid and the coherence
could therefore have an influence on the expectation values relevant
for the yield of photosynthesis only if it showed in the averaged
density matrix. The conclusion of Refs. \cite{Jiang1991a,Hoki2011a,Mancal2010a,Brumer2012a}
is that Sun light does not produce any dynamic coherence in the averaged
state of a driven photosynthetic antenna. It is less obvious how much
coherence is created when excitation is transferred into a molecular
system from a neighboring molecular system and this question will
be examined in this section. As a model for this case we take here
the chlorosome-FMO complex of the\emph{ Chlorobium tepidum} bacteria,
with the chlorosome simplified into a single pigment of its baseplate.
The FMO complex \cite{Milder2010a} is a homotrimer consisting of
bacteriochlorophyll-a (BChl-a) molecules that transfer excitation
energy between the baseplate protein of the chlorosome and the reaction
center. We study the EET dynamics considering each one of the monomers
to be independent and composed of seven BChl-a (system Hamiltonian
from Ref. \cite{Adolphs2006a}). According to the recent crystallographic
studies reporting the existence of an eighth BChl molecule \cite{Ben-Shem2004a,Tronrud2009a},
we studied the \emph{holo }form (eight BChls per monomer, Hamiltonian
from Ref. \cite{Olbrich2011a}), which presented a similar dynamics
(therefore not presented here). Based on theoretical calculations
\cite{Adolphs2006a} and recent experimental verification \cite{Wen2009a},
the FMO complex is assumed to be oriented with BChls 1 and 6 toward
the base plate protein, whereas BChls 3 and 4 are at the interface
between the FMO and the reaction center. Accordingly, the initial
excited pigments in our numerical calculations are chosen to be either
BChls 1 or 6. 

For our purpose, we postulate a simple time independent variant of
Eq. (\ref{eq:K_tdp}) with a single transfer rate from the chlorosome
baseplate to the FMO and constant rates for the internal energy relaxation
inside the FMO. For the purpose of demonstration we want to describe
correctly the relevant time-scales of the FMO's coherent oscillations,
i.e. we will use the Hamiltonian available in literature, \cite{Adolphs2006a}.
We describe energy transfer by secular Redfield tensor calculated
assuming Debye spectral density (see e.g. Eq. (3.268) in Ref. \cite{MayKuehnBook})
with a cut-off frequency $\omega_{D}=100$ cm$^{-1}$ and reorganization
energy $\lambda=$ 35 cm$^{-1}$ (as in Ref. \cite{Ishizaki2009a})
to describe the energy gap fluctuations on the FMO pigments. These
parameters give us relaxation and dephasing dynamics time-scales similar
to previous theoretical works. 

In order to quantify the coherence in the system, we define a measure
of the dynamic coherence $ $$\Xi$, as:

\begin{equation}
\Xi=\frac{\sum_{\alpha>\beta}\xi_{\alpha\beta}.(\rho_{\alpha\alpha}+\rho_{\beta\beta})}{(N-1)\sum_{1}^{N}\rho_{\alpha\alpha}}.\label{eq:acoh}
\end{equation}
Here $\xi_{\alpha\beta}=\frac{|\rho_{\alpha\beta}|^{2}}{\rho_{\alpha\alpha}\rho_{\beta\beta}}$
defines the relative amplitude of the coherences between the two excited
states $\alpha$ and $\beta$ ($\xi=1$ for a completely coherent
system, $\rho_{\alpha\beta}=\langle\alpha|\rho|\beta\rangle$). This
quantity is weighted in Eq. (\ref{eq:acoh}) by the population probabilities
of the corresponding states to only account for the significantly
populated states. It is easy to verify that $\Xi=1$ for a purely
coherent system and $\Xi=0$ when no superposition is present. Note
that the amount of coherence is a time-dependent variable when the
system of interest is subjected to relaxation due to some interaction
with bath DOF. It is a basis dependent quantity in line with the basis-dependent
properties of the dynamic coherences. Unlike established measures
such as the purity, the quantity $\Xi$, when expressed in the eigenstate
basis, is directly representative of the dynamic coherences as experimentally
accessible in the non-linear spectroscopy. When expressed in the basis
of states local to the molecules forming the aggregate, the quantity
$\Xi$ can also quantify the delocalization of a given state.

\subsection{Ultra Fast Single-Site Excitation}

Let us start with studying an ultra fast excitation of a single site
of the complex. This is an often applied but rather artificial initial
condition, as it cannot be created by light, both due to spatial (wavelength)
and spectral reasons. It is important to realize that in a system
of coupled chromophore, such as FMO, a single site represents a linear
combination of the eigenstates of the complex, and therefore its excitation
induces a coherent dynamics. This subsection discusses the established
fact that an instantaneous excitation of a spatially localized state
in an excitonically coupled aggregate induces dynamic coherence. This
fact is a basis of an assumption that the energy transfer process
in form of ``incoherent hopping'' leads on a microscopic level to
dynamic coherence on individual molecules. As we have shown in Sections
\ref{sec:Excitation-by-Neighboring} and \ref{sec:Time-Scale-of-Excitation},
even incoherent hopping has a natural time scale which is even longer
than the time scale associated with excitation by Sun-light. We have
also concluded that even if the hopping events were real, only the
ensemble average would matter. This allows us to apply certain coarse
graining in the present description of the energy transfer and to
use constant transfer, relaxation and dephasing rates.

\begin{figure*}
\includegraphics[width=1\textwidth]{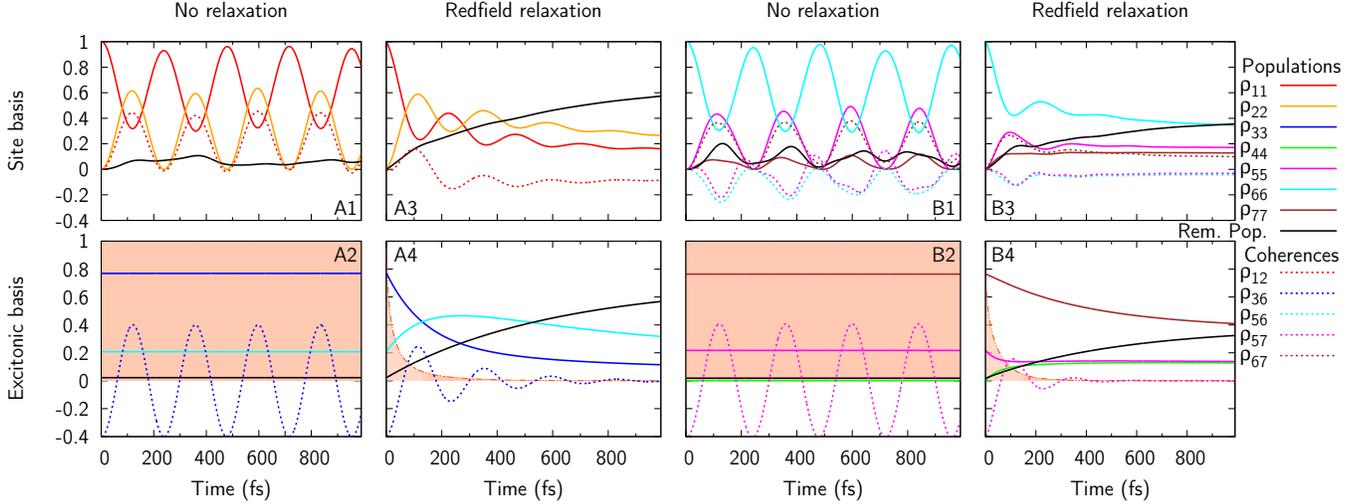}
\caption{Time evolution of populations and coherences (diagonal and off-diagonal
terms of the RDM, respectively) in the site (top) and excitonic basis
(bottom) after excitation of BChl 1 (A) or 6 (B) in a free exciton
system (1;2) or with Redfield form of relaxation (3;4). The shaded
area represents the amount of coherence in the system.}
\label{fig:7_BChl} 
\end{figure*}

Figure \ref{fig:7_BChl} presents the EET dynamics in both the site
and excitonic basis after instantaneous excitation of BChl 1 (A) or
BChl 6 (B), with or without the presence of relaxation in FMO. As
expected from the aggregate conformation, more specifically from the
excitonic heterodimer formed by BChls 1 and 2, an excitation initially
located on BChl 1 is transferred and shared with BChl 2 (see the site
basis representation, Figs. \ref{fig:7_BChl}A1 and \ref{fig:7_BChl}A3).
Populations (diagonal elements of the reduced density matrix in a
corresponding basis) display an out-of-phase evolution at the frequency
$\omega_{12}=(E_{1}-E_{2})/\hbar$, which corresponds to the frequency
of the coherence evolution. In the eigenstate basis, Figs. \ref{fig:7_BChl}A2
and \ref{fig:7_BChl}A4, the excitation of BChl 1 populates exciton
levels 3 and 6 (excitons ordered with increasing eigenenergies), which
is in concordance with previous studies (e.g. \cite{Adolphs2006a}).
We can see that the exciton populations are constant, and that only
the coherences evolve phase. Comparison of the results presented in
the site and excitonic basis (top \textit{vs} bottom) illustrates
that dynamic oscillations of the RDM off-diagonal terms, though basis
dependent, are observed in both basis. When applying a Markovian dephasing
in form of a secular relaxation tensor, beating can still be observed
on the real part of the off-diagonal terms. However, due to the interactions
with the bath, the oscillations are damped in $\sim700$ fs. This
is the time scale predicted in almost all relevant work, and its achievement
with a simple Redfield relaxation theory underlines the fact that
the details of the relaxation theory used to describe FMO, e.g. whether
Markov and secular approximations are used, is not of crucial importance.

Figure \ref{fig:7_BChl}B presents results when the excitation is
initially received by BChl 6. The relaxation pathway now involves
BChls 5 and 6 and, to a smaller extend, BChl 7 (see Figs. \ref{fig:7_BChl}B1
and \ref{fig:7_BChl}B3). Excitons 7 and 5 are the main contributors
to this transfer branch (Figs. \ref{fig:7_BChl}B2;\ref{fig:7_BChl}B4).
Similarly to the previous case, an oscillatory behavior can be observed
here, in both bases. Beating develops between sites 5 and 6 (site
basis) and between excitons 5 and 7 (excitonic basis). With the ideal
assumption of a free exciton system, undamped oscillations are observed
in both basis. When considering coupling with the bath DOF, these
oscillations are quickly damped (in $\sim400$ fs).

Note that the off-diagonal terms of the RDM, in the local basis, stabilize
around a non-zero value, but that in the basis relevant to the observation
of dynamic coherences (excitonic basis), those terms quickly equal
zero. This difference originates from the two distinct phenomena presented
in the Introduction, namely the delocalization and the dynamic coherence.
In the local basis, the non-zero values indicate delocalization of
the exciton, whereas in the excitonic basis representation, no long-lived
dynamic coherence survives when the system is interacting with its
environment.

The amount of coherence $\Xi$ (as defined in Eq. \ref{eq:acoh})
is highlighted with a shaded area to illustrate the amount of dynamic
coherence in the system. As mentioned above, this quantity is experimentally
relevant when used in the excitonic basis. In a free exciton system,
an initially coherent system remains purely coherent and accordingly
$\Xi=1$. However, the amount of coherence is damped when the system
interacts with the bath DOF, and it is verified that $\Xi(t)$ follows
a decay similar to the dynamic coherences.

\subsection{Excitation by Neighboring Antennae}

\begin{figure*}
\includegraphics[width=1\textwidth]{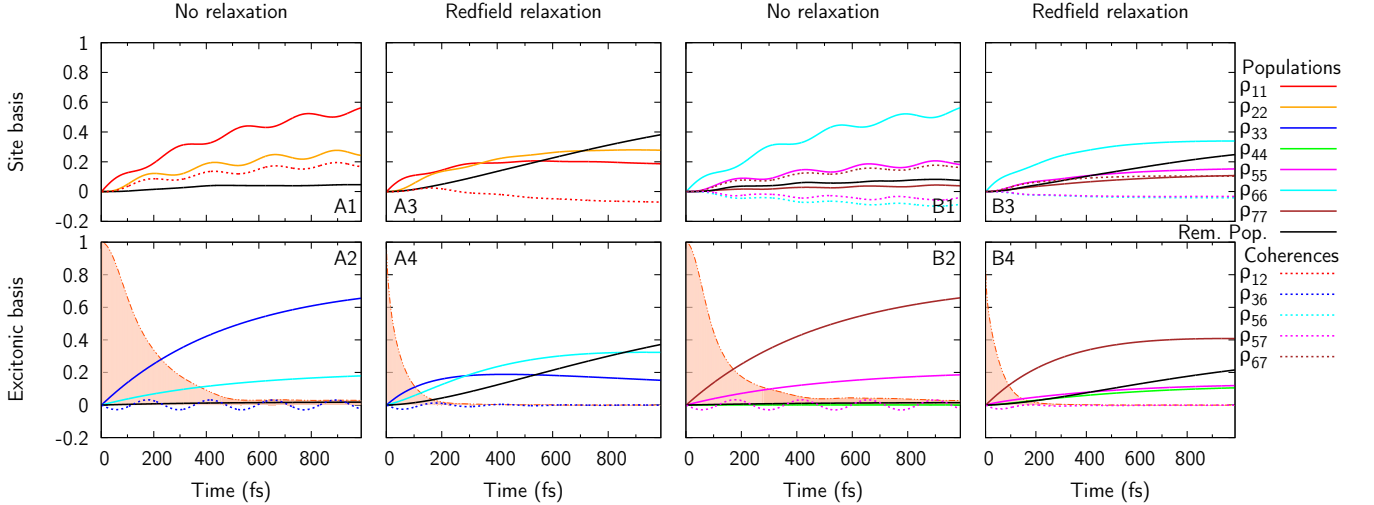}
\caption{Time evolution of populations and coherences in the site (top) and
excitonic basis (bottom) when BChl 1 (A) or 6 (B) is excited with
a feeding rate of 500 fs$^{-1}$ in a free exciton system (1;2) or
assuming a Redfield form of relaxation (3;4). The shaded area, which
represents the amount of coherence in the system, shows that even
without any relaxation dynamic coherences die out because of destructive
interferences. }
\label{fig:feed_BChl} 
\end{figure*}

Now we finally consider the response of FMO excited by its neighboring
chlorosome antennae. The complex starts in its ground state, and the
excitation is introduced from the source $s$ (representing the neighboring
BChls of the chlorosomal base plate) to the system via BChl $i$ at
a feeding rate $K_{f}$ such as: 
\begin{equation}
\frac{\partial\rho_{ii}(t)}{\partial t}=K_{f}\rho_{ss}(t).
\end{equation}
 Fig. \ref{fig:feed_BChl} shows the corresponding dynamics of EET
when BChl 1 or 6 are excited at a rate of 500 fs$^{-1}$. Results
show that the growing population of BChl 1 is slowly populating BChl
2 (see Figs. \ref{fig:feed_BChl}A1 and \ref{fig:feed_BChl}A3). In
the excitonic basis, this corresponds to populating excitons 3 and
6 (see Figs. \ref{fig:feed_BChl}A2 and \ref{fig:feed_BChl}A4). However,
even without any relaxation, this transfer does not generate a large
superposition of states, as indicated from the small amplitude of
the coherence oscillations observed in the excitonic basis (see Fig.
\ref{fig:feed_BChl}A2). Contrary to the case where a single excitation
of BChl 1 leads to a coherent dynamics (Fig. \ref{fig:7_BChl}A2),
the amount of coherence here quickly decreases with increasing populations,
and it is very low when the system is significantly populated ($\Xi<0.04$
after 500 fs$^{-1}$). This can be explained by interferences building
up between coherences of excited states created at different times
(which correspond to different phases) according to the feeding rate,
such that: $\rho_{\alpha\beta}(t)\propto\intop_{0}^{t}d\tau\, e^{i\omega_{\alpha\beta}/\hbar\, t}\, e^{i\phi(\tau)}$.
Upon continuous feeding of BChl 1 from the neighboring antenna, the
integration over different phases $\phi(\tau)$ will result in destructive
interferences, which are created within a \textsl{single} complex
and prevent any dynamic coherent superposition of excited states.

Fig. \ref{fig:feed_BChl}B shows the results when excitation from
the neighboring antennae is received by BChl 6. Such a configuration
results in a similar dynamics through the population of mainly BChl
5 and 6 and excitons 7 and 5. The evolution of the amount of coherence
is alike to the former case, without any creation of eigenstate superposition.
Relaxation through interaction with the bath DOF will certainly not
help to maintain coherent transfer, and in this case, the dynamic
coherences are damped even faster -- $\Xi(500{\rm fs}^{-1})<0.01$
in both excitation conditions -- as shown by Figs. \ref{fig:feed_BChl}A4
and \ref{fig:feed_BChl}B4.

\begin{figure*}
 \includegraphics[width=1\textwidth]{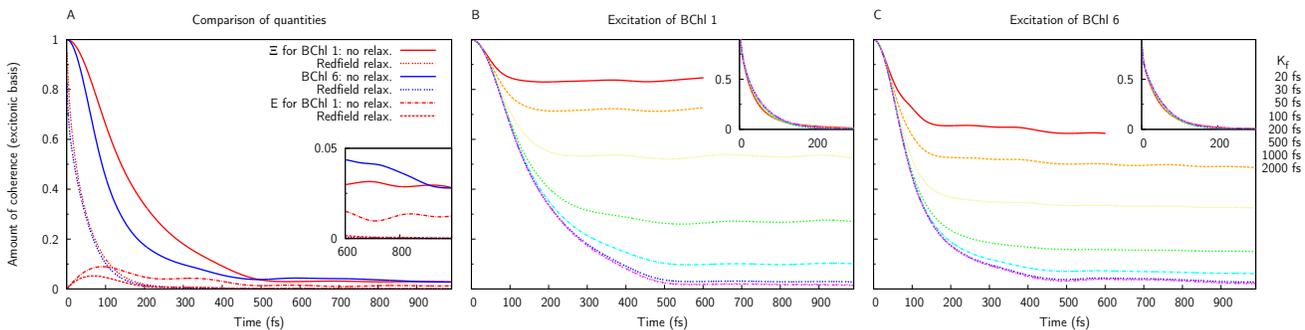}
\caption{Evolution of the amount of coherence. (A) Comparison of the amount
of coherence $\Xi$ of Eq. (\ref{eq:acoh}) and the quantity $E$
of Eq. (\ref{eq:entanglement}) with an excitation rate of 500 fs$^{-1}$
(the inset shows a detail of the results at long time). (B and C)
Amount of coherence for different excitation rates of BChl 1 and BChl
6, respectively in a free exciton system. The insets show results
using a Redfield type of relaxation. It is shown that excitation from
neighboring antennae (feeding rate of tens of picoseconds) can not
lead to a coherent superposition of excited states. \label{fig:acoh}}
\end{figure*}

Figure \ref{fig:acoh} first presents the amount of coherence calculated
according to Eq. (\ref{eq:acoh}) and comparison with the a measure
$E$ of global entanglement defined in Ref. \cite{Sarovar2010a} (Fig.
\ref{fig:acoh}A) 
\begin{equation}
E=-\sum_{\alpha}\rho_{\alpha\alpha}\ln\rho_{\alpha\alpha}-S(\rho),\label{eq:entanglement}
\end{equation}
where $S(\rho)=-{\rm tr}\{\rho\ln\rho\}$ is the von Neumann entropy
of the state $\rho$. Also we present the amount of coherence $\Xi$
for different excitation conditions (site 1 or 6 with various feeding
rates) in a free exciton system (Figs. \ref{fig:acoh}B and \ref{fig:acoh}C,
respectively) or with relaxation (insets of Figs. \ref{fig:acoh}B
and \ref{fig:acoh}C). Fig. \ref{fig:acoh}A shows that, although
the different measures of coherence provide very different values
in the system at initial time (very low populations), after a transient
effect and once the population of the excited site becomes significant
($\sim500$ fs$^{-1}$ here), both models clearly show that no dynamic
coherence survive, independently of the site excitation and dissipation
model.

Figures \ref{fig:acoh}B and \ref{fig:acoh}C show that, in a free
exciton system, the amount of coherence present in a significantly
populated system strongly depends upon the excitation rate. Independently
of the excited site (BChl 1 or 6), dynamic coherences are seen to
be created only from ultra fast excitation (faster than $\sim$100
fs$^{-1}$), i.e. when the transfer time is shorter that the period
of the coherent oscillations in the acceptor system. However, for
slower feeding rate which is representative of the natural excitation
received from the neighboring antennae (typically 1 ps$^{-1}$) \cite{Savikhin1994a,Martiskainen2009a},
the constant rate feeding of an excitation to a site does not create
any coherent superposition of excited states. As expected, coupling
to the bath generates dephasing and results in an even smaller amount
of coherence than in a free relaxation system, as shown by the insets
of Figs. \ref{fig:acoh}B and \ref{fig:acoh}C. Since excitation with
feeding rates representative of the natural excitation (tens of picoseconds)
does not lead to a superposition of coherent state already in a relaxation-free
system, it is clear that the use of the Markovian dynamics does not
limit our results. This is in agreement with Ref. \cite{Jesenko2013a}.
In other words, a system interacting with a slowly relaxing bath supporting
non-Markovian evolution of excitons would, under excitation from neighboring
antennae, also not lead to a coherent dynamics, which is in contrast
with the concern recently raised in Ref. \cite{Fassioli2012a}.

In our demonstration calculation we have not treated non-secular effects.
Recent experimental observations suggest that non-secular effects,
through the coupling of population with coherence terms, could generate
dynamic coherences \cite{Panitchayangkoon2010a}. It has been analytically
shown \cite{Pachon2012b} that non-secular effect could indeed be
responsible for coherent dynamics but only in the very short transient
interval time following excitation. Here, even if coherences would
be created, the continuous excitation from the neighboring antennae
would lead to destructive interferences and suppress any coherent
superposition.

\section{Conclusions\label{sec:Conclusions}}

In this paper we clarify the difference between the two effects often
related to as quantum coherence. We suggest that only the eigenstate
delocalization influences the yield and efficiency of excited state
energy transfer in molecular aggregates, while the dynamic quantum
coherence is an artifact of the particular laboratory excitation conditions.
We find that in classical systems one can naturally view the slow
external driving process as an integrated action of individual sudden
excitations. One can represent a real classical system by a virtual
ensemble of identical systems experiencing these sudden excitations,
and it is likely that physical representations of such virtual ensembles
may exist for some class of problems. 

When attempting a similar representation of a quantum mechanical dynamics
of an open molecular system interacting weakly with light or a neighboring
molecular system, we find that a representation by sudden excitations
is not satisfactory for field with finite spectral width and for neighboring
antennae with finite emission spectrum width. Although the spectral
width of the light defines a corresponding natural time scale of the
excitation event, it is not possible to decompose the interaction
with the field or a neighboring system into excitation events by some
unique way. Because the yield of photosynthesis is an intrinsically
averaged quantity, one can make an almost trivial observation that,
in second-order response description, no matter how the quantum expectation
value is decomposed, whether into some average-like contributions
of individual antennae or into contributions exhibiting dynamic coherence
on a microscopic level, the average remains the same. The dynamic
coherence would have to be present in this average in order to be
of relevance. We propose a simple experiment which could in principle
test the relevance of the short time-scale of excitation events.

For a model system with important parameters taken from the most frequently
studied molecular antenna exhibiting long living coherence, the FMO
complex, we demonstrate that under its natural timescale of excitation
no significant presence of dynamic coherence can be found, even if
the system itself could avoid all the dephasing induced by its protein
environment. Although we believe that the description of the excitation
process represented by the density matrix is correct even on the level
of individual molecules, without any subdivision of the excitation
fields, the contrary would nevertheless lead to the same result for
the efficiency of photosynthesis, and to the same conclusion, namely,
that the dynamic coherence is not relevant for the yield of photosynthesis.
The time evolution of the dynamic coherence generated by short excitation
in photosynthetic systems can provide us with invaluable information
about the internal dynamics of the photosynthetic energy transfer,
and it can thus serve as an important diagnostic tool. However, the
suggested significance of the dynamic coherence for natural light-harvesting
cannot be currently experimentally confirmed, nor can it be theoretically
expected based on t quantum mechanics.

\bibliographystyle{prsty}

\end{document}